\pdfoutput=1
\documentclass[reprint, pra]{revtex4-2}

\usepackage{amsthm}
\usepackage{amsmath}
\usepackage{latexsym}
\usepackage{amsfonts}
\usepackage{amssymb}
\usepackage{color}
\usepackage{bbm,dsfont}
\usepackage{graphicx}
\usepackage{enumerate}
\usepackage{hyperref}
\usepackage{subfigure} 
\usepackage[dvipsnames]{xcolor}
\usepackage{tikz}
\usepackage{placeins}
\usepackage{import}
\usepackage{enumitem}
\usepackage[mathscr]{eucal}

\usepackage[english]{babel}
\usepackage[utf8]{inputenc}
\newcommand\opone{\leavevmode\hbox{\small1\kern-3.8pt\normalsize1}}

\DeclareUnicodeCharacter{00A0}{ }

\usepackage{bbold}
\usepackage[normalem]{ulem}

\newcommand{\W}{\mathscr{W}}
\newcommand{\U}{\mathscr{U}}

\newcommand{\Y}{\mathscr{Y}}
\newcommand{\Tr}{\operatorname{Tr}}
\newcommand{\T}{\mathscr{T}}

\newcommand{\sigmap}{\operatorname{\sigma_+}}
\newcommand{\sigmam}{\operatorname{\sigma_-}}

\newtheorem{proposition?}{Proposition?}

\theoremstyle{definition}

\definecolor{ao(english)}{rgb}{0.0, 0.5, 0.0}
\definecolor{darkorange}{RGB}{255, 126, 0.0}
\definecolor{denim}{rgb}{0.08, 0.38, 0.74}
\definecolor{burntorange}{RGB}{204, 85, 0}

\newcommand{\id}{\mathbbm{1}} 

\newcommand{\Sys}{\mathcal{S}} 
\newcommand{\Sysone}{{\mathcal{S}_1}} 
\newcommand{\Systwo}{{\mathcal{S}_2}} 
\newcommand{\Mem}{\mathcal{M}} 
\newcommand{\Memone}{{\mathcal{M}_1}} 
\newcommand{\Memtwo}{{\mathcal{M}_2}} 
\newcommand{\Bath}{\mathcal{B}} 

\newcommand{\Qdot}{|\dot{Q}|} 
\newcommand{\Qcritmin}{|\dot{Q}|^{\textrm{crit}}_\textrm{min} } 
\newcommand{\Qcritmax}{|\dot{Q}|^{\textrm{crit}}_\textrm{max} } 
\newcommand{\Qmin}{\big|\dot{Q}\big|_\textrm{min}} 
\newcommand{\Qmax}{\big|\dot{Q}\big|_\textrm{max} } 
\newcommand{\Cmax}{C_\textrm{max}}
\newcommand{\sigint}{\sigma_\textrm{int}}

\usepackage{physics}
\usepackage{appendix}
\usepackage{epstopdf}

\begin{document}


\title{Quantum memory enhanced dissipative entanglement creation in non-equilibrium steady states}

\author{Daniel Heineken} \email{daniel.heineken@studium.uni-hamburg.de}
\affiliation{Institut f{\"u}r Theoretische Physik, Technische
  Universit{\"a}t Dresden, D-01062, Dresden, Germany}

\author{Konstantin Beyer} \email{konstantin.beyer@tu-dresden.de}
\affiliation{Institut f{\"u}r Theoretische Physik, Technische
  Universit{\"a}t Dresden, D-01062, Dresden, Germany}

\author{Kimmo Luoma} \email{kimmo.luoma@tu-dresden.de}
\affiliation{Institut f{\"u}r Theoretische Physik, Technische
  Universit{\"a}t Dresden, D-01062, Dresden, Germany}

\author{Walter
  T. Strunz} \email{walter.strunz@tu-dresden.de} \affiliation{Institut
  f{\"u}r Theoretische Physik, Technische Universit{\"a}t Dresden,
  D-01062, Dresden, Germany}

\date{2021-03-02}
     
\begin{abstract}
    This Article investigates dissipative preparation of entangled non-equilibrium steady states (NESS).
    We construct a collision model where the open system consists of two qubits which are coupled to
    heat reservoirs with different temperatures. The baths are modeled by sequences of qubits
    interacting with the open system. The model can be studied in different
    dynamical regimes: with and without environmental memory effects. We report that only a certain bath temperature range allows for entangled NESS. Furthermore, we obtain minimal and maximal critical values for the heat current through the system.
    Surprisingly, quantum memory effects play a crucial role in 
    the long time limit. First, memory effects broaden the parameter region where quantum correlated NESS may be dissipatively prepared and, secondly, they increase the attainable concurrence. Most remarkably, we find a heat current range that does not only allow, but even guarantees that the NESS is entangled. Thus, the heat current can witness entanglement of non-equilibrium steady states.
\end{abstract}


\maketitle

\section{Introduction}
Entanglement is one of the key resources for quantum information processing and quantum
technologies~\cite{Ac_n_2018}. It is known to be a fragile property of multipartite quantum states which 
is easily lost due to thermal fluctuations or decoherence emerging from the unavoidable 
coupling of the system of interest to external degrees of freedom. One may try to 
counteract the dissipative effecs to protect the fragile quantum properties~\cite{PhysRevLett.105.200402,PhysRevLett.111.030405} 
or utilize them as a part of the state preparation procedure ~\cite{PlenioCavitylossinducedGenerationEntangled1999,ArnesenNaturalThermalMagnetic2001,SchneiderEntanglementSteadyState2002,PhysRevLett.106.090502,KrausPreparationEntangledStates2008, VerstraeteQuantumComputationQuantumstate2009,Lin2013,ReiterSteadystateEntanglementTwo2013,ShankarAutonomouslyStabilizedEntanglement2013,WalterEntanglementNanoelectromechanicalOscillators2013,TavakoliHeraldedGenerationMaximal2018,romanancheyta2020enhanced}. 

In this Article, we investigate quantum correlations in dissipatively prepared non-equilibrium steady states (NESS). While the field of equilibrium quantum thermodynamics is well established~\cite{gemmer2004quantum,binder2019thermodynamics}, non-equilibrium thermodynamics is rapidly developing. Concepts such as quantum fluctuation theorems~\cite{CrooksQuantumoperationtime2008,RasteginNonequilibriumequalitiesunital2013,AlhambraFluctuatingWorkQuantum2016a,AbergFullyQuantumFluctuation2018,DebarbaWorkestimationwork2019a,RibeiroExperimentalstudygeneralized2020}, thermodynamic uncertainty relations~\cite{PhysRevLett.114.158101,FalascoUnifyingthermodynamicuncertainty2020,Horowitz2020,HasegawaQuantumThermodynamicUncertainty2020}, and quantum heat engines~\cite{Skrzypczyksmallestrefrigeratorscan2011,UzdinEquivalenceQuantumHeat2015,FriedenbergerWhenquantumheat2017,ThomasThermodynamicsnonMarkovianreservoirs2018,DasQuantumenhancedfinitetimeOtto2020,CamatiEmployingnonMarkovianeffects2020,Hovhannisyan_2019} but also the basic definitions of work and heat in quantum systems and the fundamental differences from their classical counterparts are still under debate~\cite{Gooldrolequantuminformation2016,StrasbergQuantumInformationThermodynamics2017,Perarnau-LlobetCollectiveoperationscan2019,BeyerSteeringHeatEngines2019,Garcia-PintosFluctuationsExtractableWork2020,Roman_Ancheyta_2019}. 

One way to prepare a non-equilibrium steady state is to couple an open system to two heat reservoirs
with differing temperatures and letting the open system relax. Due to the temperature difference of the heat reservoirs, a typical characteristic of the NESS is a persistent heat current through the  open system~\cite{BermudezControllingMeasuringQuantum2013,CharalambousHeatcurrentcontrol2019,MedinaGonzalezHeatflowreversaltrappedion2020}.
We investigate such a scenario by coupling an open quantum system, consisting of two qubits, to a hot and a cold thermal reservoir. We use collision models, which have become a very popular tool to analyze open quantum system dynamics in recent years~\cite{collision_models_ciccarello,AltamiranoUnitarityfeedbackinteractions2017,PhysRevA.99.042108,GuarnieriNonequilibriumsteadystatesmemoryless2020a,SeahNonequilibriumDynamicsFinitetime2019,RodriguesThermodynamicsWeaklyCoherent2019}. They provide a transparent approach for describing the dynamics of quantum correlations between the open 
system and its environment~\cite{Pellegrini_2009,FilippovDivisibilityquantumdynamical2017,BeyerCollisionmodelapproachsteering2018}, which is particularly interesting for the understanding of non-Markovian quantum dynamics~\cite{kretschmer_collision_2016,CampbellSystemenvironmentcorrelationsMarkovian2018,10.1007/978-3-030-31146-9_3}.  Another growing field of applications for collision models are quantum transport phenomena, which also include energy transport due to heat currents~\cite{LorenzoHeatfluxquantum2015,LiEffectcoherencenonthermal2018,MicadeiReversingdirectionheat2019,KhandelwalCriticalheatcurrent2020a,De_Chiara_2018}.

In this Article, we especially study how memory effects in the relaxation dynamics influence the quantum correlations of the NESS. 
In general, the occurrence of entanglement in the NESS is restricted to certain temperature regions of the heat baths and requires a critical minimal heat current, as reported in~\cite{KhandelwalCriticalheatcurrent2020a}. However, in our model also a maximal critical value for the heat current exists, beyond which the NESS is necessarily separable again. We find that memory effects increase these parameter regions, thus, allowing for a build-up of entanglement for temperatures and heat currents which would {\it always} lead to a separable NESS in the memoryless case. This allows to certify memory effects in the relaxation dynamics from the steady state properties of the system. 
Analyzing the relation between the heat current and the maximal possible concurrence in the NESS, we observe a further surprising feature emerging from the memory effects. Namely, certain heat current values do not only allow, but even {\it guarantee} that the corresponding NESS is entangled.

The outline of the remainder of this Article is the following. In Sec.~\ref{sec:coll_model} we describe our collision model. Then in Sec.~\ref{sec:memoryless_case} we analyze the memoryless case. Memory effects are included in Sec.~\ref{sec:memory_dynamics} before we study the relation between entanglement and heat current in Sec.~\ref{sec:heat-current}. In Sec.~\ref{sec:non-markovian} we comment on the non-Markovianity of the quantum dynamics generated by the model.  We present our final conclusions in Sec.~\ref{sec:conclusion}.

\section{Collision model}\label{sec:coll_model}

\begin{figure}[ht]
  \centering
  \includegraphics{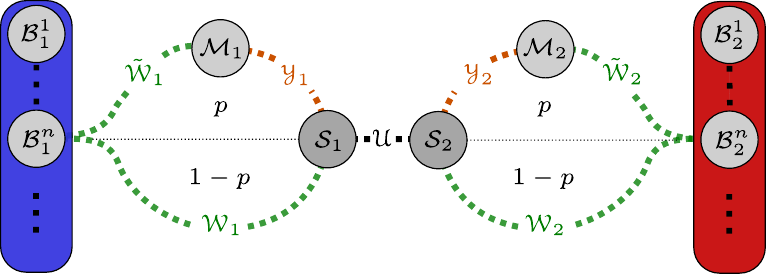}
  \caption{Sketch of the collision model. The two-qubit system $\Sysone+\Systwo$ couples to a sequence of hot and cold reservoir subenvironments ($\Bath_{1,2})$. Dotted lines show possible interactions. The interaction between $\Sys$ and $\Bath$ can be either direct (with probability $(1-p)$) or mediated by memory qubits $\Mem_{1,2}$ (with probability $p$).}
    \label{fig:systemsketch_nonmark}
\end{figure}

In this work, we investigate a two-qubit system $\Sys= \Sys_1 + \Sys_2$ with
Hamiltonian $H_\Sys = \frac{\omega}{2}(\sigma_z^{\Sys_1} + \sigma_z^{\Sys_2})$ which couples
to two thermal reservoirs $\mathcal{B}_1$ and $\mathcal{B}_2$ with
temperatures $T_1$ and $T_2$. In the framework of collision models,
the reservoirs are modeled as products of qubit subenvironments where each subenvironment interacts only once with the system (see
Fig.~\ref{fig:systemsketch_nonmark}). 
 By $\Bath_{1,2}^n$ we denote the $n$th
subenvironment in $\Bath_1$ or $\Bath_2$, respectively. 
 The local Hamiltonian for each
subenvironmental qubit is given by
$H_{\Bath^n_{1,2}} = \frac{\omega}{2}\sigma_z$. The subenvironments are initially in a thermal state $\xi_{1,2} = \frac{1}{2}(\mathds{1} + z_{1,2}\,\sigma_z)$, where $z_{1,2} \in [-1, 0]$ are temperature parameters related to the Boltzmann factor of the respective temperatures $T_1$ and $T_2$ \cite{PhysRevLett.88.097905}:
  \begin{equation}
    z = \frac{1-e^{\frac{\omega}{k_BT}}}{1+e^{\frac{\omega}{k_BT}}}.
  \end{equation}
  A value of $z=-1$ corresponds to $T=0 \, \mathrm{K}$, while $z = 0$ is equivalent to  $T \rightarrow \infty$. Thus, the state of each subenvironment before the collision is given by
  \begin{equation}
    \xi = \xi_1 \otimes \xi_2 = \frac{1}{2}(\mathds{1} +
    z_1\,\sigma_z) \otimes \frac{1}{2}(\mathds{1} + z_2\, \sigma_z).
  \end{equation}
For $z_{1,2}\in(0,1]$ the subenvironment qubit is in a population inverted state and we call such a state an {\it inverted thermal state} in this Article. 

We additionally introduce two memory qubits $\Memone$ and $\Memtwo$ which are not discarded in between the collisions and, thus, allow information to propagate over the sequence of system-bath interactions. 
Each collision model step consists of an inner-system transformation mediated by $\U$, which is followed by an interaction between each of the system qubits with its respective bath. With a probability $(1-p)$, the subenvironmental qubit $\Bath_i^n$ couples directly via $\W_i$ to the system qubit $\Sys_i$. With a probability $p$, $\Bath_i^n$ couples via $\tilde{\W}_i$ to the respective memory qubit $\Mem_i$, which then interacts via the operation $\mathscr{Y}_i$ with $\Sys_i$.
Thus, the model effectively describes a situation where each system qubit is damped by two baths, one with and one without memory. The parameter $p$ can tune the relative weight between these different damping channels.
For $p=1$, this model is equivalent to the collision model presented in Ref.~\cite{kretschmer_collision_2016}, where system environment correlations are propagated by swapping the "collided" and "fresh" subenvironment in between system-environment interactions. In the context of dissipative entanglement generation a similar model has been studied in Ref.~\cite{ZnidaricEntanglementStationaryNonequilibrium2012, znidaric_2010}. For 
$p>0$, the collision model is closely related to the models presented in Refs.~\cite{PhysRevA.87.040103,PhysRevA.93.052111}, where the swapping is done probabilistically.

The interactions are given by the excitation number conserving unitary operators
\begin{align}
  \begin{split}
    &\U = \U_{\Sysone \Systwo} = e^{-i \Omega \Delta t[\sigma_+ \otimes \sigma_{-} \,+\, \sigma_-\otimes\sigma_+]}, \\
    &\W_{i} = \W_{\Bath_i\Sys_i} = e^{-i \sqrt{\Gamma_i\Delta t}[\sigma_+ \otimes \sigma_{-} \,+\, \sigma_-\otimes\sigma_+]}, \\
    &\tilde{\W}_{i} = \tilde{\W}_{\Bath_i\Mem_i} = e^{-i \sqrt{\Gamma_i \Delta t} [\sigma_+ \otimes \sigma_{-} \,+\, \sigma_-\otimes\sigma_+]},\\
    &\Y_{i} = \Y_{\Sys_i\Mem_i} = e^{-i \Upsilon_i\Delta t[\sigma_+ \otimes \sigma_{-} \,+\, \sigma_-\otimes\sigma_+]}, \\
    \label{eq:operators}
  \end{split}
\end{align}
where $\Omega, \Gamma_{1,2}, \Upsilon_{1,2}$ are the respective coupling strengths and $\Delta t$ is the duration of one collision.
The probability that one of the two system qubits couples via the memory qubit in a given collision model step is independent from the type of interaction that is undergone by the other qubit.
We, thus, have four possible one-step maps for the evolution of the joint state $\rho_{\Sys\Mem}$ of the system and the memory qubits.
We define the operators
 \begin{align}
 \label{eq:T-operators}
  \begin{aligned}
   &\T_{1} = \tilde{\W_1}\Y_1\tilde{\W_2}\Y_2\U,\\       &\T_{3} = \tilde{\W_1}\Y_1\W_2\U,
  \end{aligned}
  &&
  \begin{aligned}
   &\T_{2}= \W_1\W_2\U, \\       &\T_{4}=\W_1\tilde{\W}_2\Y_2\U,
  \end{aligned}
 \end{align}
where $\T_1$ describes the coupling via both memory qubits, $\T_2$ the direct coupling between system and the reservoirs and the $\T_{3,4}$ describe the scenarios, in which one system qubit couples directly to the reservoir while the other one couples via its memory qubit.
The four one-step maps are, thus, given by
\begin{equation}
  \mathcal{E}_i[\rho_{\Sys\Mem}] = \mathrm{Tr}_\Bath[\T_i(\rho_{\Sys\Mem} \otimes \xi)\T^\dagger_i].
\end{equation}
Therefore, the one step map for the evolution of $\rho_{\Sys\Mem}$ is the statistical mixture
\begin{gather}
\label{eq:discrete_map}
  \rho^{n+1}_{\Sys\Mem} = \mathcal{E}[\rho^n_{\Sys\Mem}],\\
      \mathcal{E} = p^2\mathcal{E}_{1}+ (1-p)^2\mathcal{E}_{2} + p(1-p)[\mathcal{E}_{3}+ \mathcal{E}_{4}].
\end{gather}
The model is, by construction, discrete in time. However,
the different scaling in $\Delta t$ in Eq.~\eqref{eq:operators} is chosen such that a
time-continuous limit of the dynamics can be derived~\cite{collision_models_ciccarello}.
By expanding the map in
Eq. \eqref{eq:discrete_map} up to first order in $\Delta t$ and taking
the limit $\Delta t \rightarrow 0$, we obtain the following 4-qubit Gorini-Kossakowski-Sudarshan-Lindblad (GKSL) master equation governing the dynamics of system and memory together~\cite{doi:10.1063/1.522979,lindblad1976}:
\begin{align}
    \label{eq:4-qubit-GKSL}
    \dot\rho_{\Sys\Mem}&(t) = -i [\Omega \sigint^{\Sysone\Systwo} 
    +  p \sum_{k=1}^2 \Upsilon_k \sigint^{\Sys_k\Mem_k}, \rho_{\Sys\Mem}(t)] \nonumber \\
    +& (1-p) \sum_{k=1}^2 \Gamma_k \big( z_k^- \mathcal{D}[\sigmam^{\Sys_k}] + z_k^+ \mathcal{D}[\sigmap^{\Sys_k}] \big)\rho_{\Sys\Mem}(t)\nonumber\\
    +& p \sum_{k=1}^2 \Gamma_k \big( z_k^- \mathcal{D}[\sigmam^{\Mem_k}] + z_k^+ \mathcal{D}[\sigmap^{\Mem_k}] \big)\rho_{\Sys\Mem}(t)\notag\\
    &=\mathcal{L}_{\Sys\Mem}\left[\rho_{\Sys\Mem}(t)\right].
\end{align}
with  $z^\pm_{k} = (1 \pm z_{k})/2$, $\mathcal{D}[\mathrm{A}]\rho = \mathrm{A}\rho\mathrm{A}^\dagger - \frac{1}{2}(\mathrm{A}^\dagger \mathrm{A} \rho + \rho \mathrm{A}^\dagger \mathrm{A})$, $\sigint = (\sigmap \otimes \sigmam + \sigmam \otimes \sigmap)$ and the superscripts indicate the subsystems on which the operators act. We note that, in accordance with the fermionic nature of a qubit bath, the rates $z^\pm_{k}$ agree with the dissipation rates for a fermionic reservoir, i.e., $z^+ = n_F$ and $z^- = 1 - n_F$, where $n_F$ is the Fermi-Dirac distribution. 

\section{Memoryless Case}\label{sec:memoryless_case}
We first consider the memoryless scenario \mbox{($p=0$)}, i.e., the system qubits always couple directly to the reservoir subenvironments by the unitary transformations $\W_1$ and $\W_2$ and the memory qubits $\Memone$ and $\Memtwo$ are not involved in the interaction. A similar model has been studied earlier, for example, in  Refs.~\cite{BraskAutonomousquantumthermal2015,KhandelwalCriticalheatcurrent2020a}.


The dynamics of the system alone is given by the GKSL master equation
\begin{align}
  \begin{split}
    \dot{\rho}_\Sys(t) = &-i\Omega[\sigma_+ \otimes \sigma_- + \sigma_- \otimes \sigma_+, \rho_\Sys(t)] \\
 &+\Gamma_1 (z^-_1 \mathcal{D}[\sigma_-\otimes\mathds{1}] +  z^+_1 \mathcal{D}[\sigma_+\otimes\mathds{1}])\rho_\Sys(t) \\
&+\Gamma_2 (z^-_2 \mathcal{D}[\mathds{1} \otimes \sigma_-] +  z^+_2 \mathcal{D}[\mathds{1} \otimes \sigma_+])\rho_\Sys(t)
  \end{split}\notag\\
=&\mathcal{L}\left[\rho_\Sys(t)\right].
  \label{eq:master_eq}
\end{align}
As a consequence of Eq. \eqref{eq:master_eq}, the evolution of the system reads
\begin{equation}
  \rho_\Sys(t) = e^{(t-t_0) \, \mathcal{L}}[\rho_\Sys(t_0)],
\end{equation}
with the generator $\mathcal{L}$.
For the further treatment of the problem, we introduce the ratios $\gamma_{1,2}$ between the system-bath and the inner-system coupling:
\begin{equation}
  \gamma_{1,2} = \Gamma_{1,2}/{\Omega}.
\end{equation}

\subsection{Steady state} \label{subsec:steady_state_markov}
The steady state $\rho^\infty_\Sys$ of the open system satisfies
\begin{equation}
  \mathcal{L
  }[\rho^\infty_\Sys]=0.
\end{equation}
It follows directly from the form of the generator $\mathcal{L}$ in Eq.~\eqref{eq:master_eq}, that this steady state only depends on the two ratios $\gamma_{1,2}$.
For $\gamma_{1,2} >0$, the system has a unique steady state $\rho^\infty_\Sys(z_{1,2}, \gamma_{1,2})$ which is reached in the limit $t \rightarrow \infty$.
This steady state can be written as
\begin{equation}
  \rho^\infty_\Sys =  \rho^\infty_\Sysone \otimes  \rho^\infty_\Systwo+\chi,
  \label{eq:rho_ss_markov}
\end{equation}
with the local reduced states of the two qubits $\rho^\infty_{\Sys_{1,2}}$ and the matrix
\begin{equation}
  \chi =
  \begin{pmatrix}
    -\eta^2  & 0 & 0 & 0 \\
    0 & \eta^2 & i\eta & 0 \\
    0 & -i\eta & \eta^2 & 0 \\
    0 & 0 & 0 &-\eta^2
  \end{pmatrix},
  \label{eq:chi}
\end{equation}
with the real valued function
\begin{equation}
  \eta = (z_1-z_2)\frac{\gamma_1\gamma_2}{(\gamma_1+\gamma_2)(\gamma_1\gamma_2+4)}.
  \label{eq:eta_markov}
\end{equation}
$\chi$ describes the correlations arising between the two subsystems in the computational basis.
We find that in the steady state regime the local states of the two system qubits are thermal states themselves:
\begin{equation}
  \rho^{\infty}_{\Sys_{1,2}} = \mathrm{Tr}_{\Sys_{2,1}}[\rho^{\infty}_\Sys] = \frac{1}{2}(\mathds{1}+s_{{1,2}}\,\sigma_z).
\end{equation}
The parameters $s_{{1,2}}$ given by
\begin{equation}
    s_1 = z_1 - 4\frac{\eta}{\gamma_1} , \quad
    s_2 = z_2 + 4\frac{\eta}{\gamma_2},
\end{equation}
describe the temperatures of the individual qubits $\Sysone$ and $\Systwo$.
For $z_1 \lessgtr z_2$ (i.e., $T_1 \lessgtr T_2$) a temperature gradient can be observed:
\begin{equation}
  z_1 \lessgtr s_{1} \lessgtr s_{2} \lessgtr z_2.
\end{equation}
As can be directly seen from the equations above, for $z_1 = z_2$ no correlations build up and the steady state is just the thermalized product state
\begin{equation}
  \rho^\infty_\Sys(z_1=z_2) = \xi_1\otimes\xi_1 = \xi_2\otimes \xi_2.
\end{equation}
Another limit is reached for $\gamma_1, \gamma_2 \gg 1$ (i.e., the system-bath coupling is much stronger than the inner-system coupling), in which case we observe $\eta \rightarrow 0$ and
\begin{equation}
  \rho^\infty_\Sys(\gamma_1, \gamma_2 \gg 1) \approx \xi_1\otimes\xi_2.
  \label{eq:high_gamma_limit}
\end{equation}
Although this might be an expected result, it interestingly still holds if just one of the two system-bath couplings is chosen to be much stronger than the inner-system coupling, in which case not just the strongly coupled qubit almost thermalizes with its bath, but also the state of the 
second system qubit shifts towards the thermal state of the subenvironment it couples to.

\subsection{Heat current}
Due to the coupling of the two-qubit open system to thermal
reservoirs, heat is transferred between the baths and the
system. As pointed out in Ref.~\cite{PhysRevE.97.022111}, the heat exchanged between the system and the reservoirs during one collision can be seen as the energy change in
the subenvironment qubits taking part in the interaction.

We note that defining the heat current as an observable on the bath qubits is conceptually different from a heat current definition based on the dissipators of the master equation. While being quantitatively equivalent, the latter case, in general, would not be an observable. For the remainder of this Article a measurable heat current is, however, crucial in order to establish a connection with the entanglement in the NESS. Therefore, we make use of the collision model, which transparently includes the baths, to introduce the heat current as follows.

With the Hamiltonian $H_{\Bath_1^n}$ for the $n$-th qubit in $\Bath_1$ and the operator $\T_2$ in Eq.~(\ref{eq:T-operators}), the change of energy in $\mathcal{B}_1$ during the $n$-th interaction with the system is given by
\begin{equation}
    \Delta E_1^n = \Tr[H_{\Bath_1^n}(\Lambda^n[\xi_1^n] - \xi_1^n)],
\end{equation}
with
\begin{equation}
  \Lambda^n[\xi_1^n] = \Tr_{\Sys, \Bath_2}[\T_2(\rho^{n-1}_\Sys \otimes \xi^n)\T_2^\dagger]
\end{equation}
being the quantum channel for the $n$th environmental qubit interacting with the system.
When the system has reached the steady state $\rho_\Sys^\infty$ we 
find in the time-continuous limit
\begin{align}
  \begin{split}
    \lim_{\Delta t \to 0} \frac{\Delta E_1^\infty}{\Delta t} = \dot{\widetilde Q}_1 &= -2\,\frac{ (z_1-z_2)\, \omega \Omega \, \gamma_1\gamma_2}{(\gamma_1+\gamma_2)(\gamma_1\gamma_2 + 4)} \\
    &= -2 \, \omega \Omega \, \eta.
  \end{split}
  \label{eq:heat_flow_markov_1}
\end{align}
Analogously, we find for $\Bath_2$
\begin{equation}
  \dot{\widetilde Q}_2= -\dot{\widetilde Q}_1.
  \label{eq:heat_flow_markov_2}
\end{equation}
Throughout the rest of the Article we analyze the scaled heat current
\begin{align}
    \dot Q = \frac{\dot{\widetilde Q}}{\omega \Omega}.
\end{align}

From Eqns. \eqref{eq:heat_flow_markov_1} and \eqref{eq:heat_flow_markov_2} we can immediately see that the heat current follows the temperature gradient between the two baths (i.e.,  $\dot{Q}_1 \lessgtr 0$ for $T_1 \gtrless T_2$) and grows with increasing temperature difference between the two baths as we would expect.
The largest heat current between the two baths is achieved for $\gamma_1 = \gamma_2 = 2$, i.e., the case in which the coupling between the system qubits and the respective reservoirs is twice as strong as the inner-system coupling:
\begin{equation}
  \Qmax = \bigg|\frac{1}{4}\, (z_1-z_2)\bigg|.
  \label{eq:markov_max_heat}
\end{equation}

In analogy to Sec.~\ref{subsec:steady_state_markov}, we can also observe that for $\gamma_1 \gg 1$ the value of  $\Qmax$ decreases as $1/\gamma_{1}$, leading to a heat insulating effect for the two qubit system. The same result holds for $\gamma_2 \gg 1$.

\subsection{Steady state entanglement of the system} \label{conc_markov}
The matrix $\chi$ in Sec.~\ref{subsec:steady_state_markov} describes the correlations between the two open system qubits. Naturally, the question arises under which circumstances the two qubits are entangled in the steady state.

\begin{figure}[ht]
  \centering
  \includegraphics{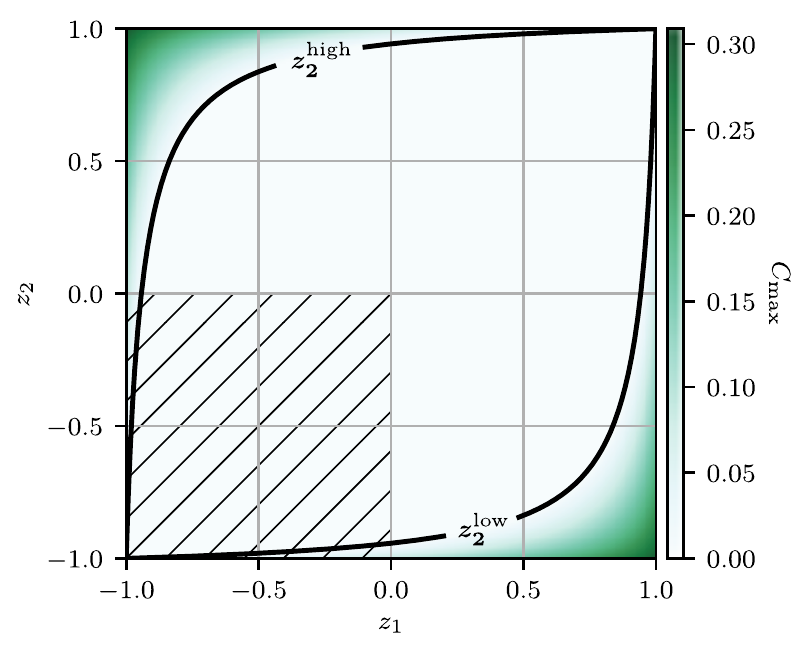}
  \caption{Largest possible concurrence in the steady state for given $z_{1,2}$. The black line corresponds to the boundary for pairs $(z_1, z_2)$ for which $\Cmax=0$. For each pair of bath temperatures $z_1$ and $z_2$ the coupling parameters are numerically optimized to achieve the maximum concurrence. The hatched area highlights the region in which the bath qubits $\xi_1, \xi_2$ are in thermal states.}
  \label{fig:concurrence_markov_lim}
\end{figure}

As can be seen from Eqns.~\eqref{eq:rho_ss_markov} and \eqref{eq:chi}, the steady state of the two-qubit system is an \textit{X-state} with $\rho_{14}=\rho_{41}=0$. Thus, the concurrence of the steady state is given by \cite{Yu:2007:EED:2011832.2011835}
\begin{equation}
  C = 2\cdot \max\{0,|\rho_{23}| - \sqrt{\rho_{11}\rho_{44}}\},
  \label{eq:conc_mark}
\end{equation}
and is a function of the temperatures $z_{1,2}$ and the coupling parameters
$\gamma_{1,2}$. In order to investigate which bath temperatures lead to an
entangled NESS, we numerically optimize $\gamma_{1,2}$ to find the maximum concurrence
$C_{\max}$ for given values $z_{1,2}$. The results
are visualized in Fig. \ref{fig:concurrence_markov_lim}.
The hatched area corresponds
to the region in which the environmental baths are in thermal states ($z_{1,2} \in [-1, 0]$), while in the non-hatched area at least one of the reservoirs consists of qubits in inverted thermal
states (with $z_i \in (0, 1]$). 

We find that steady state entanglement can be created for all pairs $(z_1, z_2)$ satisfying
\begin{equation}
  z_1 z_2 + \sqrt{\frac{9}{8}}\,|z_1 - z_2| >1.
\end{equation}
Thus, above (below) the boundary 
\begin{equation}
  z_2^{\text{high,low}}= \frac{4\pm 3\sqrt{2}\,z_1}{4\,z_1 \pm 3\sqrt{2}}
\end{equation}
the values of $\gamma_1$ and $\gamma_2$ can be chosen such that the NESS is entangled (see Fig.~\ref{fig:concurrence_markov_lim}). The parameters yielding the boundary  $C_{\max} = |\rho_{23}| - \sqrt{\rho_{11}\rho_{44}}=0$ are given by
\begin{equation}
  \gamma_1^{\text{high,low}} = \frac{2}{\sqrt{2}\pm z_1},\, \gamma_2^{\text{high,low}} = 4\sqrt{2}-\gamma_1^{\text{high,low}}.
\end{equation}
In the next Section we will see that memory effects in the relaxation dynamics can increase the correlations in the NESS.




\section{Coupling with Memory}\label{sec:memory_dynamics}
In what follows, the memory qubits $\Mem_{1,2}$ are taken into account in the collision model ($p>0$). We will investigate to what extent dynamical memory effects have an influence on the entanglement in the non-equilibrium steady state of our heat transport model. As shown earlier, entanglement in the NESS is only possible in a certain temperature regime. As we will see, memory effects can increase the parameter region where entanglement can occur.

For $p>0$ the dynamics of the open system, in general, cannot  be described by a GKSL master equation but could be obtained from the 4-qubit evolution of $\Sys$ and $\Mem$ by tracing out the memory qubits
\begin{align}
    \rho_\Sys(t) = \Tr_\Mem[\rho_{\Sys\Mem}(t)].
\end{align}
Here, we are only interested in the steady state. Therefore, we compute the steady state $\rho_{\Sys\Mem}^\infty$ of the 4-qubit GKSL dynamics (Eq.~(\ref{eq:4-qubit-GKSL})) and recover the system steady state as $\rho_\Sys^\infty = \Tr_\Mem[\rho_{\Sys\Mem}^\infty]$. A closed analytical solution cannot be given for an arbitrary choice of the model parameters and, thus, the results for the case with memory have been evaluated numerically.



\subsection{Steady state entanglement}
\label{sec:steady-state-entanglement}

\begin{figure}[htp]
	\centering
	\includegraphics{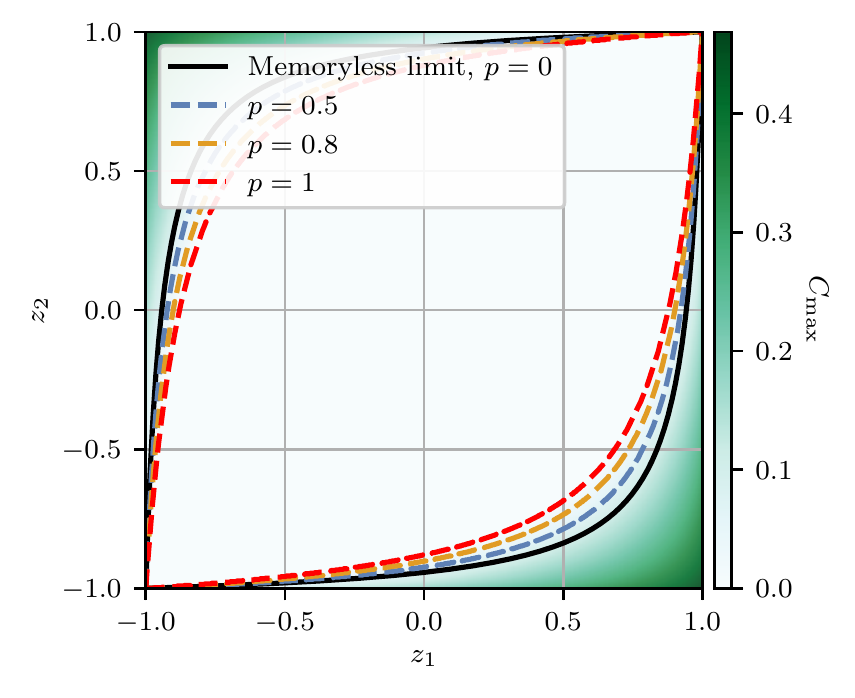}
	\caption{The dashed lines indicate how the boundary of the region where entangled NESS are feasible changes with the memory parameter $p$. The density plot of the maximum concurrence $\Cmax$ corresponds to the case of full memory \mbox{$p=1$}. $\Cmax$ is obtained by numerically optimizing all coupling parameters in the model for the given temperatures $(z_1,z_2)$.}
	\label{fig:concurrence_nonmarkov}
\end{figure}

In analogy to the approach in Sec.~\ref{conc_markov}, we analyze the entanglement that can be generated between the two qubits $\Sys_1$ and $\Sys_2$ in the steady state regime. By optimizing the coupling parameters
\begin{align}
\gamma_{1,2} = \Gamma_{1,2}/\Omega, && \upsilon_{1,2} =\Upsilon_{1,2}/\Omega,
\end{align}
to maximize the concurrence between $\Sys_1$ and $\Sys_2$ for a fixed pair of environmental temperature parameters $(z_1, z_2)$, we obtain Fig.~\ref{fig:concurrence_nonmarkov}. We observe that the region of pairs $(z_1, z_2)$ in which entanglement can be created ($\Cmax>0$) monotonically increases with $p$. The value of the concurrence achievable for a fixed pair $(z_1, z_2)$ is higher than in the case of the memoryless coupling (Fig.~\ref{fig:concurrence_markov_lim}) considered before. In fact, for all pairs $(z_1, z_2)$, the reachable entanglement $\Cmax$ between the two system qubits grows also monotonically with $p$ and is maximal for $p=1$. Thus, the entanglement in the steady state can witness memory effects in the environment if the concurrence reaches values beyond the limit attainable by the memoryless scenario.

\section{Relation between heat current and entanglement}
\label{sec:heat-current}
We have seen a clear connection between the reachable steady state entanglement $\Cmax$ and the memory parameter $p$, always with respect to a fixed pair of bath temperatures $z_{1,2}$. Even though this gives insight for which temperatures one can expect to find entanglement in the steady state at all, the correct temperature regime is, of course, not sufficient. For each pair of temperatures one can always find coupling parameters which lead to a separable steady state (e.g., by setting $\gamma_{1,2} \rightarrow \infty$). 

\begin{figure}[ht]
	\centering
	\includegraphics[width=\columnwidth]{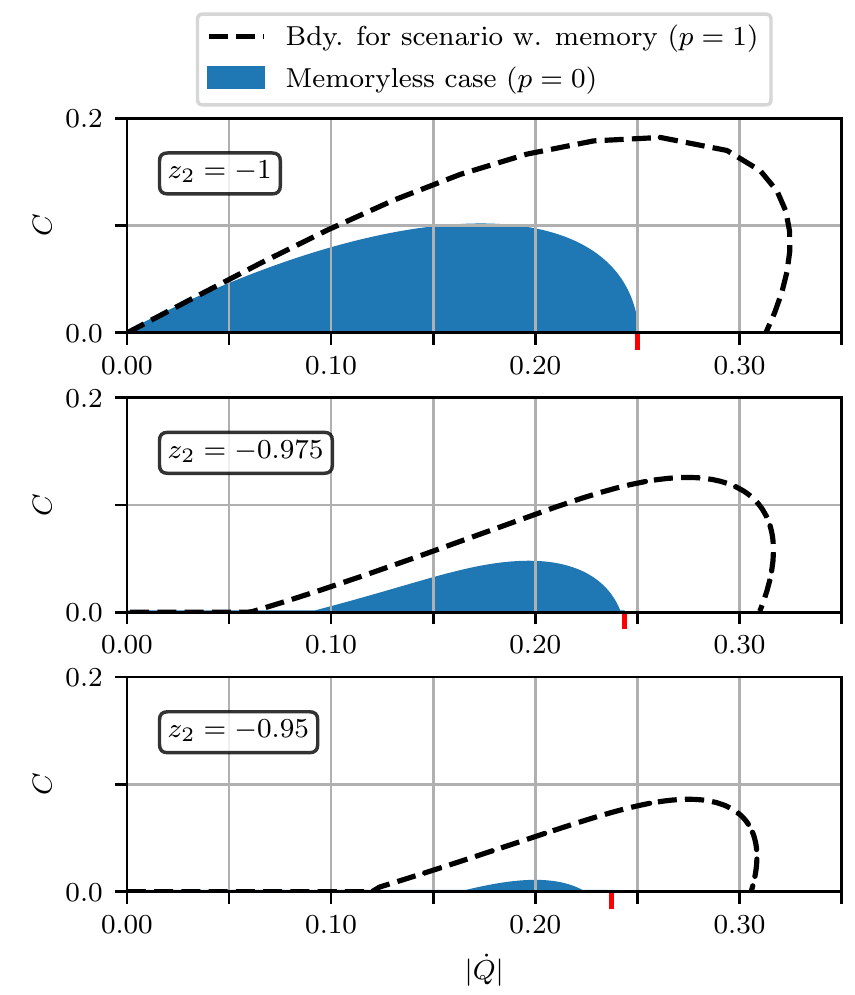}
	\caption{The blue area depicts all possible pairs $(\vert\dot{Q}\vert, C)$ of steady state heat current vs. concurrence for a fixed pair of temperatures in the memoryless scenario. The dashed black line shows the boundary of the area in the case with memory ($p = 1$) for the respective temperatures. In all plots, $z_1 = 0$, while $z_2 \in\{-1, -0.975, -0.95\}$ (from top to bottom). The red tick on the abscissa denotes the maximal heat current $\vert \dot{Q}\vert_\mathrm{max}$ which can be obtained for the respective choice of temperatures in the memoryless scenario (cf. Eq.~\eqref{eq:markov_max_heat}). It is important to note that there are always states with $C=0$ for all $0 \leq \vert\dot{Q}\vert \leq \vert \dot{Q}\vert_\mathrm{max}$.}
	\label{fig:markov_heat_entanglement_thermal}
\end{figure}

Therefore, in order to get a better understanding of how the energy transport influences the quantum correlations in the NESS, we will now investigate the relation between the reachable steady state entanglement and the heat current through the system. 
The definition of the heat current is independent of the type of coupling between the baths and, therefore, allows to compare the cases with and without memory. Moreover, in both cases all interactions are excitation preserving. This ensures that the different behaviour between the two cases does not rely on some external energy input hidden in the different interactions. 

For the memoryless case, it follows from Eqns.~\eqref{eq:conc_mark},~\eqref{eq:heat_flow_markov_1} and the form of the steady state given by Eq.~\eqref{eq:rho_ss_markov}, that the steady state is entangled if and only if \begin{equation}
    \vert \dot{Q} \vert > 2 \sqrt{\rho_{11}\rho_{44}}.
    \label{eq:crit_heat}
\end{equation}
This condition was firstly introduced in Ref.~\cite{KhandelwalCriticalheatcurrent2020a}. It provides a necessary and sufficient criterion for steady state entanglement, provided one has full knowledge of the temperatures and coupling parameters. In particular, entangled steady states require a non-zero heat flow. In this section, we analyze relations between heat flow and entanglement that can be observed without any knowledge of the coupling parameters of the system. 

\subsection{Critical heat current}

To understand which amount of 
concurrence $C$ can be obtained for a given heat current $\Qdot$, we scan the steady states for a large set of coupling parameters $\gamma_{1,2}\in (0, 1000]$ and for fixed temperatures $z_1=0$ and $z_2 \in \{-1,-0.975,-0.95\}$. The solid blue area in Fig.~\ref{fig:markov_heat_entanglement_thermal} shows which concurrences can be obtained in a memoryless scenario. For comparison, the dashed line gives the boundary for the case with maximal memory ($p=1$). As we might expect already from previous considerations, the memory effects enlarge the region that supports NESS which are entangled. 

\begin{figure}[ht]
	\centering
	\includegraphics[width=\columnwidth]{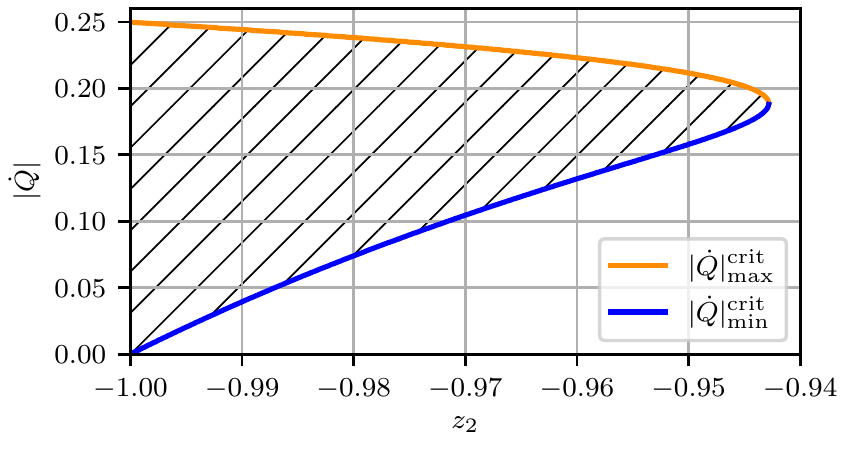}
	\caption{Critical heat current values as a function of $z_2$ for fixed $z_1 = 0$ in the memoryless case. The hatching depicts the region for which we can potentially find entangled steady states.}
	\label{fig:crit_heat_flow}
\end{figure}

For a given pair of temperatures, there are critical values $\Qcritmin$ and  $\Qcritmax$ that constrain the heat current for which entanglement can form. In general, these thresholds  differ from the extremal heat currents $\Qmin = 0$ and $\Qmax$ (see Eq.~(\ref{eq:markov_max_heat})). In Fig.~\ref{fig:crit_heat_flow} we plot the critical values for the memoryless case in dependence of the temperature $z_2$ for a fixed temperature $z_1 = 0$. It is worth noting, that the condition
\begin{equation}
    \Qcritmin < \vert \dot{Q} \vert < \Qcritmax
\end{equation}
establishes a necessary, but not sufficient criterion for steady state entanglement. However, the boundaries depend solely on the bath temperatures and not on the coupling parameters.

\subsection{Heat current as a witness for entanglement}
\begin{figure}[ht]
	\centering
	\includegraphics[width=\columnwidth]{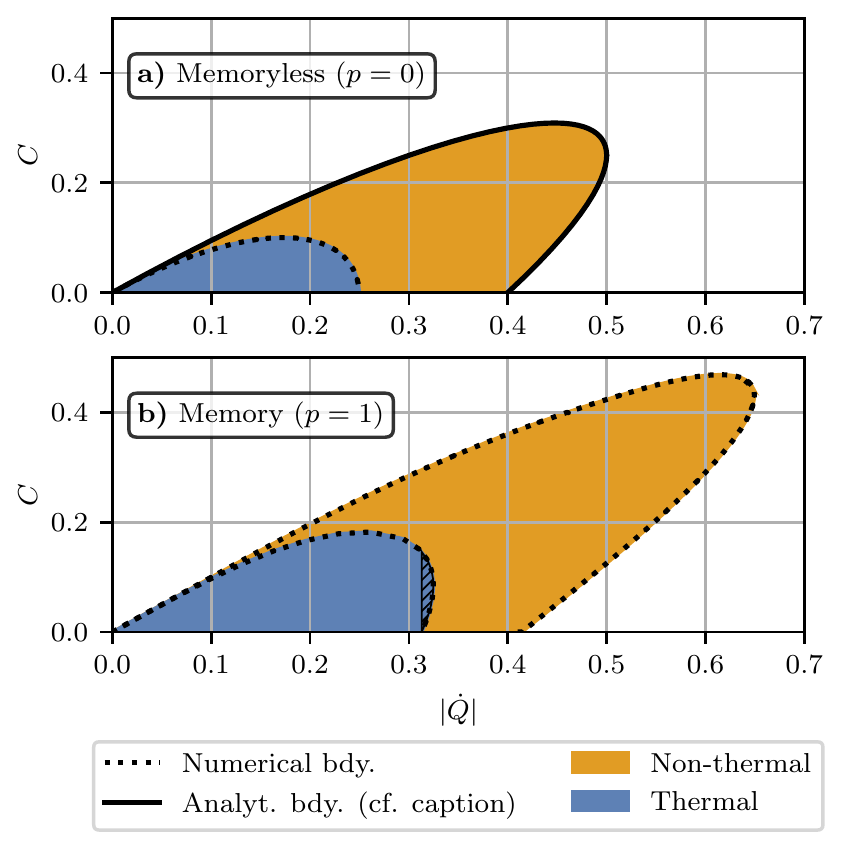}
	\caption{$C(\vert \dot{Q} \vert)$ areas for a) the memoryless scenario and b) coupling with memory. The blue area corresponds to the $(\vert \dot{Q} \vert, C)$-values that can be reached by steady states when coupled to thermal environments ($z_1 = 0, z_2 = -1$), while the orange area depicts the points that are obtained when the system couples to a thermal and an inverted thermal environment ($z_1 = 1, z_2 = -1$). The hatching in b) highlights the region where an overhang exists in the thermal regime. The corresponding heat current interval ensures entanglement in the system. The solid black boundary is given by $\Qdot= \frac{1}{5}(1 + 4C \mp \sqrt{1-2C-4C^2})$. The dotted black lines are numerical boundaries obtained by finding the extremal values of $C$ for a fixed $\vert \dot{Q}\vert$.}
	\label{fig:c_q_tot}
\end{figure}

The area in the $C(\Qdot)$ plot (Fig.~\ref{fig:markov_heat_entanglement_thermal}) depends on the choice of the temperatures $z_{1,2}$ and is maximized for the largest possible temperature difference. In  Fig.~\ref{fig:c_q_tot} we plot $C(\Qdot)$ again for the maximal thermal temperature difference ($z_1 = -1$, $z_2 = 0$) in comparison to the non-thermal case  ($z_1 = -1$, $z_2 = +1$) for couplings with and without memory. The non-thermal curve in Fig.~\ref{fig:c_q_tot} a) (memoryless scenario) shows an overhang. Thus, there are heat current values which can only be obtained if the corresponding non-equilibrium steady state is entangled. However, it has to be noted that the second bath is not in a thermal but in an inverted thermal state.

Fig.~\ref{fig:c_q_tot} b), where we consider the coupling with memory ($p=1$), shows interesting details. The curves look similar to the memoryless case but the heat current region for which entanglement is possible increases and the reachable concurrence is higher, as could be expected already from the results of Sec.~\ref{sec:memoryless_case}. Most remarkably, the overhang now also shows up for the thermal case (both heat reservoirs in thermal states). This means that there is a certain range for the heat current which guarantees that the steady state of the system is entangled.
An experimentally measurable heat current could, therefore, witness entanglement in the system if this suitable interval is reached.
The blue area in Fig.~\ref{fig:c_q_tot} b) only shows the case of a maximal thermal temperature spread, where the overhang becomes most visible. However, the respective $C(\vert \dot{Q} \vert)$ areas for smaller temperature differences always lie inside the boundary of the maximal case, ensuring that a heat current in the overhang range is an entanglement witness independently of the concrete choice of temperatures.

Crucially, the heat current can only serve as an entanglement witness if it can be measured by an observable. This is the case in our collision model setup which explicitly includes the baths. In a master equation approach, solely defined on the system, the heat current is in general not an observable and, therefore, could not constitute such a witness.

\section{On the divisibility of the dynamics}
\label{sec:non-markovian}
Memory effects are often related to a non-Markovian behavior of the open quantum system dynamics. Even though we mainly study the influence of memory effects on the steady state properties of the system, we would like to comment on the non-Markovianity of the quantum dynamics that lead to these steady states. In particular we provide numerical evidence that the steady states with the maximal entanglement for given temperatures are always reached by non-divisible quantum dynamics.

Non-Markovianity is usually studied as a transient phenomenon. Several nonequivalent criteria for non-Markovian quantum dynamics have been proposed in the literature~\cite{BreuerColloquiumNonMarkovianDynamics2016,HuelgaNonMarkovianityAssistedSteadyState2012,BreuerMeasureDegreeNonMarkovian2009}. We will focus here on the geometrical P-divisibility criterion introduced in Ref.~\cite{lorenzoGeometricalCharacterizationNonMarkovianity2013} because of its numerical stability for the given data.

A quantum dynamical map $\Lambda_t: \rho(0) \rightarrow \rho(t) = \Lambda_t[\rho(0)]$ is called \textit{CP-divisible} if it can be decomposed as
\begin{align}
\label{eq:lambda-decomposition}
    \Lambda_t = \Lambda_{t,s} \Lambda_s,
\end{align}
where the two-times map $\Lambda_{t,s} = \Lambda_t \Lambda^{-1}_s$ is completely positive and trace-preserving (CPT) for all $s<t$. If $\Lambda_{t,s}$ is positive but not completely positive for some values $t,s$ then the dynamics is said to be \textit{P-divisible}. A dynamical map $\Lambda_t$ is $\textit{indivisible}$ if it is not P-divisible.

P-divisibility is characterized by a monotonic decrease of the state space volume reachable by the dynamics~\cite{lorenzoGeometricalCharacterizationNonMarkovianity2013}.
A density operator $\rho$ of a $d$-dimensional quantum system can be decomposed as
\begin{align}
    \rho = \sum_{i=0}^{d^2-1} \Tr[\rho \,G_i]\,G_i = \sum_{i=0}^{d^2-1} \vec{r}_i \,G_i,
\end{align}
where the $G_{1,\ldots, d^2-1}$ are the Hermitian and traceless generators of the group $SU(d)$ and $G_0 = \id/\sqrt{d}$. The vector $\vec{r}$ can be seen as a generalized Bloch vector.
The map $\Lambda_t$ can then be written in the basis $\{G_i\}$, acting on $\vec{r}$ as
\begin{align}
    \vec{r}(t) = F(t)\,\vec{r}(0), && F_{i j}(t) = \Tr[G_i\, \Lambda_t [G_j]],
\end{align}
and the concatenation of maps in Eq.~(\ref{eq:lambda-decomposition}) is given by the matrix multiplication
\begin{align}
    F(t) = F(t,s)\, F(s).
\end{align}
It has been shown in Ref.~\cite{WolfDividingQuantumChannels2008} that the absolute value of the determinant of a positive map $F(s)$ can only decrease under composition with another positive map $F(t,s)$: 
\begin{align}
    |\det F(s)| \geq |\det \left(F(t,s)\, F(s)\right)| = |\det F(t)|.
\end{align}
Thus, if the dynamics is P-divisible (i.e., $F(t,s)$ is positive for all $t>s$) then $|\det F(t)|$ can only decrease over time. An increase of $|\det F(t)|$ indicates that the dynamics is not P-divisible and, therefore, also not CP-divisible.

Based on this observation, the following measure for non-divisibility has been proposed in Ref.~\cite{lorenzoGeometricalCharacterizationNonMarkovianity2013}:
\begin{align}
\label{eq:non-divisibility}
    \mathcal{N} = \int\displaylimits_{\frac{\partial}{\partial t} |\det F(t)| > 0}\; \frac{\partial}{\partial t} |\det F(t)|\, dt.
\end{align}

\begin{figure}[ht]
	\centering
	\includegraphics[width=\columnwidth]{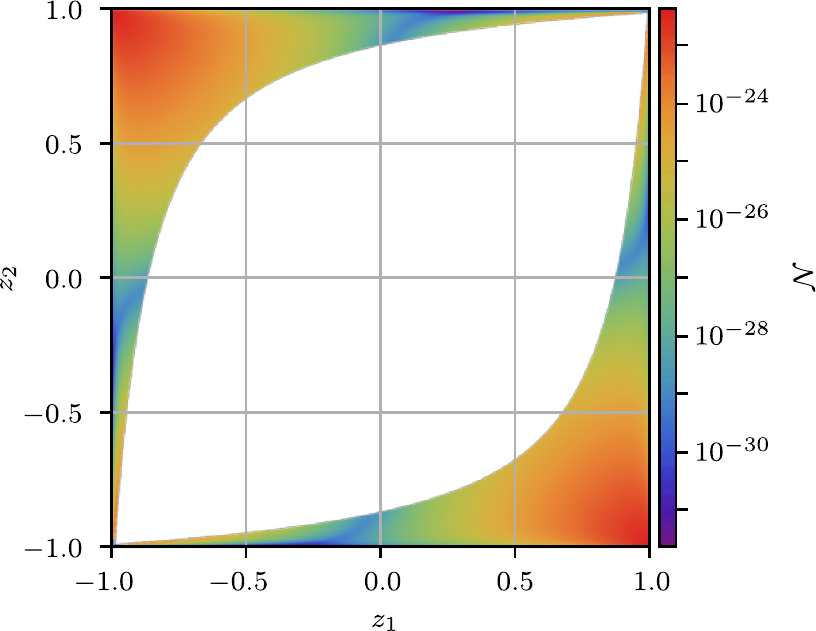}
	\caption{For the optimized coupling parameters that lead to the highest steady state entanglement $\Cmax$ for a given temperature pair $(z_1,z_2)$ we plot the non-divisibility $\mathcal{N}$ of the associated quantum dynamics in the memory case $p=1$. The dynamics which lead to the maximal entanglement $\Cmax$ in the NESS are always non-divisible. However, the non-divisibility $\mathcal{N}$ is not monotonically related to the maximal concurrence $\Cmax$ as can been seen by a comparison with Fig.~\ref{fig:concurrence_nonmarkov}. The spread of $\mathcal{N}$ over several orders of magnitude is related to the fact that $|\det F(t)|$ typically decays exponentially over time~\cite{lorenzoGeometricalCharacterizationNonMarkovianity2013}. The magnitude of $\mathcal{N}$ then strongly depends on where the time range with $\frac{\partial}{\partial t} |\det F(t)| > 0$ is situated with respect to the time scale of the decaying dynamics which, in turn, depends on the interplay of the different coupling parameters in the model.}
	\label{fig:divisibility}
\end{figure}

Using the non-divisibility measure $\mathcal{N}$ we numerically analyze the dynamics generated by our model with memory:
\begin{align}
    \Lambda_t[\rho] = \Tr_\Mem\left[e^{t\,\mathcal{L_{\Sys\Mem}}}[\xi_1 \otimes \rho \otimes \xi_2] \right],
\end{align}
with the generator $\mathcal{L}_{\Sys\Mem}$ as defined in Eq.~(\ref{eq:4-qubit-GKSL}).

First of all one should note that a collision model with memory effects (i.e., $p>0$ in our setup) does not guarantee non-divisible dynamics. Even in the case $p=1$ the resulting dynamics can be divisible depending on the choice of the coupling parameters. This is an expected behavior and well described in the literature~\cite{kretschmer_collision_2016,CarusoQuantumChannelsMemory2014}.

To make the connection to the main question of this paper, namely the entanglement generation in the steady state, we focus here on the dynamics which induce the maximal possible concurrence $\Cmax$ in the NESS for given temperatures $(z_1, z_2)$. We can numerically verify that the optimized parameters found in Sec.~\ref{sec:steady-state-entanglement}, i. e, those that generate maximal entanglement $\Cmax$, always lead to $\mathcal{N} > 0$. In other words, the maximal steady state entanglement is reached through indivisible dynamics.

In Fig.~\ref{fig:divisibility} we show $\mathcal{N}$ for the temperature region where the NESS of the dynamics is entangled ($\Cmax>0$). While the temperature pairs with maximal non-divisibility $(z_1 = \pm 1, z_2 = \mp 1)$ agree with those that reach the maximal $\Cmax$, there is in general no monotonic relation between $\mathcal{N}$ and $\Cmax$ as one can see by comparing Fig.~\ref{fig:divisibility} to Fig.~\ref{fig:concurrence_nonmarkov}. In particular, the maximal thermal temperature spread $(z_1=-1, z_2=0)$ shows a rather low $\mathcal{N}$ whereas the associated $\Cmax$ at this point is maximal for the thermal regime.

We can state that our model is able to generate non-Markovian quantum dynamics and that maximal entanglement in the NESS even requires the dynamics to be indivisible. However, a clear relation between transient non-Markovianity and steady state quantities such as quantum correlations in the NESS is missing to date and this interesting question needs further investigation.

\section{Conclusions}
\label{sec:conclusion}
In this Article we investigate the heat flow through a two-qubit open quantum system which is in contact with two heat reservoirs at different temperatures. Using a collision model approach, we consider different types of couplings between the system and the baths, implementing dynamics with and without memory effects.

The focus of our work lies on the entanglement content of non-equilibrium steady states of the open quantum system. We show that entanglement can only persist for a certain range of reservoir temperatures. Memory effects in the relaxation dynamics increase the temperature range allowing for entanglement. Thus, steady state entanglement can build up for temperature pairs which would {\it always} lead to a separable steady state in a memoryless scenario. 
Accordingly, the occurrence of entanglement for those temperatures serves as a witness for memory effects in the relaxation dynamics.

The non-equilibrium steady state, its entanglement, and the heat current strongly depend on the concrete choices for the several coupling parameters in the model. For any pair of temperatures one can find coupling parameters which lead to a separable NESS. However, maximum entanglement and heat current are closely related as we show in the second part of this Article. For given temperatures, a critical minimum heat current can be obtained which is necessary to allow steady state entanglement at all. Interestingly, in general, there is also an upper critical heat current beyond which the corresponding steady state is always separable.  

Memory effects again broaden the range between the lower and upper bound. Additionally, a surprising effect becomes visible in this scenario. For heat current values close to the upper critical limit, steady state entanglement is not only possible but even necessary. Thus, observing suitable heat currents in such systems guarantees an entangled steady state irrespective of any details about the couplings involved.  

The dynamics which lead to maximal entanglement in the NESS are always indivisible. Thus, for this special case we can establish a connection between a steady state property and a dynamical characteristics. However, further investigation is needed to gain more general insights.

Our work shows that memory effects, which are often studied rather in the context of dynamical phenomena, can play an important role for the build-up of quantum correlations in non-equilibrium steady states. Especially the heat current interval which ensures entanglement can be interesting for dissipative preparation of entangled states.

\acknowledgments{The authors would like to thank Francesco Ciccarello for illuminating discussions.}
\bibliographystyle{unsrtnat}
\bibliography{bibliography}

\end{document}